\documentclass[prx,twocolumn,aps,
nolongbibliography,floatfix]{revtex4-1}

\usepackage[dvipsnames]{xcolor}
\usepackage{tikz}
\usepackage{graphicx}
\usepackage{dcolumn}
\usepackage{amsfonts}
\usepackage{amsmath,amssymb,amsthm}
\usepackage{bm}
\usepackage{bbm}
\usepackage{dsfont}
\usepackage{physics}
\usepackage{multirow}
\usepackage{hyperref}
\usepackage{cleveref}
\usepackage{xcolor}
\usepackage{placeins}
\usepackage[normalem]{ulem}
\usepackage{MnSymbol,comment}
\usepackage[english]{babel}

\newcommand{\kk}{{\bm k}}

\newcommand{\Heff}{H_{\mathrm{eff}}}

\newcommand{\HALM}{H_{\mathrm{ALM}}}

\newcommand{\lamind}{\lambda_R^{\mathrm{ind}}}   
\newcommand{\DALM}{\Delta_{\mathrm{ALM}}}        
\newcommand{\Dv}{\Delta_v}                       
\newcommand{\Dc}{\Delta_c}                       
\newcommand{\xal}{x}                             
\newcommand{\etav}{\eta_v^{0}}                   
\newcommand{\etac}{\eta_c^{0}}                   
\newcommand{\gv}{g_v}
\newcommand{\gc}{g_c}

\newcommand{\bk}{{\bm k}}

\newcommand{\cS}{{\cal S}}
\newcommand{\cT}{{\cal T}}

\newcommand{\cH}{{\cal H}}

\theoremstyle{definition}

\usetikzlibrary{arrows.meta,decorations.markings,patterns}

\begin{document}
\title{Engineering nonrelativistic spin splitting in semiconductors by altermagnetic proximity}
\date{\today}

\author{Sing Lam~Wong}
\email{slwongag@student.ubc.ca}
\affiliation{Department of Physics and Astronomy, University of British Columbia, Vancouver, British Columbia V6T 1Z1, Canada}
\affiliation{Quantum Matter Institute, University of British Columbia, Vancouver, British Columbia V6T 1Z4, Canada}

\author{Marcel~Franz}

\affiliation{Department of Physics and Astronomy, University of British Columbia, Vancouver, British Columbia V6T 1Z1, Canada}
\affiliation{Quantum Matter Institute, University of British Columbia, Vancouver, British Columbia V6T 1Z4, Canada}

\begin{abstract}
Altermagnets combine vanishing net magnetization with momentum-dependent, nonrelativistic spin splitting, offering functionalities attractive for spintronics and quantum devices. Their practical use, however, is constrained by the limited availability of materials that simultaneously possess the required altermagnetic symmetry, semiconducting electronic structure, and weak spin-orbit coupling. Here we show that these requirements can instead be separated between constituent materials through proximity engineering. We develop an analytical theory for a conventional semiconductor coupled to a gapped $d$-wave altermagnet and demonstrate that virtual tunneling transfers the altermagnetic spin splitting to the semiconductor. Crucially, the induced nonrelativistic and relativistic spin splittings exhibit qualitatively different dependences on band alignment. Virtual processes involving the valence and conduction bands of the altermagnet can reinforce the altermagnetic splitting while contributing with opposite signs to the interfacial Rashba coupling. Consequently, electrostatic tuning can access a regime in which the Rashba splitting vanishes while a sizable $d$-wave spin splitting persists. Exact diagonalization of the full heterostructure model confirms the analytical theory and establishes the robustness of this mechanism beyond the perturbative regime. Our results identify altermagnetic proximity as a route for engineering nonrelativistic spin-split electronic states in established semiconductor platforms, providing a general strategy for disentangling magnetic and relativistic spin interactions in quantum heterostructures.

\end{abstract}

\date{\today}
\maketitle

\section{Introduction}\label{sec:intro}

\begin{figure}[t]
\includegraphics[width=\columnwidth]{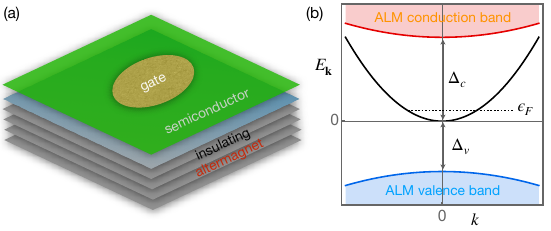}
\caption{The proposed setup. (a) A thin semiconductor film is
  supported on an insulating altermagnet. Electrostatic top gate can
  be used to define an elliptical quantum dot that can function as a
  field-
  free spin qubit. (b) Band structure: conduction band of the
  semiconductor (solid black line) is assumed to lie inside the gap of
  the insulating altermagnet. The band alignment is described by two
  parameters $\Delta_v$ and $\Delta_c$ that measure the position of
  the semiconductor band edge with respect to the ALM valence and
  conduction band, respectively.  }
\label{fig1}
\end{figure}

Assembling quantum materials with disparate physical properties into
heterostructures can enable devices with novel
or highly improved functionalities. Notable examples of this paradigm
include the
interface between two insulating perovskite oxides LaAlO$_3$ and SrTiO$_3$
which forms a high-mobility electron gas \cite{Hwang2004}, Weyl
semimetals that can be engineered in thin film multilayers of a
magnetically doped 3D topological insulator separated by ordinary-insulator spacer layers \cite{Burkov2011}, 
 and topological superconductors in one and two spatial dimensions
 that can be formed by combining various existing ingredients
 \cite{Fu2008,Sau2010,Alicea2010,Oreg2010,Lutchyn2010,Cao2018,Can2021}. Here
 we employ this time-tested strategy to engineer an {\em altermagnetic
 semiconductor} by interfacing an ordinary non-magnetic semiconductor
 with an insulating altermagnet, as schematically depicted in Fig.\
 \ref{fig1}. As we discuss below semiconductors with altermagnetic
 properties could be technologically valuable in spintronics and qubit design but no single material
 is currently known to be suitable for such applications.

 Altermagnets are a new class of collinear magnetically ordered
materials  whose magnetic sublattices are related by rotation rather than
translation or inversion
\cite{Smejkal2022a,Smejkal2022b,Mazin2022,Jungwirth2026}. Their
characteristic signature is a net zero magnetization accompanied by momentum-dependent, 
single-electron band  spin splitting that can have $d$-, $g$- or
$i$-wave symmetry in the Brillouin zone, depending on the
material. MnTe is a canonical example of a semiconducting altermagnet
\cite{Mazin2023,Krempasky2024,Amin2024} 
while CrSb is an altermagnetic metal \cite{Ding2024,Reimers2024}. Importantly, both are $g$-wave
altermagnets which limits their technological usefulness. The most
interesting applications, such as the spin-splitter effect
\cite{Smejkal2021,Zarzuela2025} and its
superconducting counterpart \cite{Monkman2026}, have been predicted to
occur in $d$-wave altermagnets and they are symmetry-forbidden in $g$-
and $i$-wave materials.

The goal, therefore, is to engineer a $d$-wave altermagnetic
semiconductor with suitable properties from material ingredients that are
available. In support of this goal we consider here  a minimal model of 
an insulating $d$-wave altermagnet, henceforth  referred to as the ``donor",
brought into contact with a thin-film single-valley semiconductor, the
``host", in a heterostructure shown schematically in Fig.\
 \ref{fig1}(a). Our  model calculations show that significant
 spin-splitting can be induced in the semiconducting host through the
 altermagnetic proximity effect.

 The underlying microscopic mechanism relies on
 virtual tunneling of electrons from the host to the donor and back,
 which produces $d$-wave altermagnetic splitting in the former proportional to the
 second power of the tunneling amplitude. 
 For realistic material
 parameters, the energy scale governing this splitting, which we denote $\eta^{\rm ind}_0$, can be significant, typically tens of meV,
 reflecting values of spin splitting in the host altermagnet that can
 be as large as 1 eV. The actual spin splitting experienced by electrons near the Fermi surface depends on the carrier density in the semiconductor through its dependence on the Fermi momentum $\sim \eta^{\rm ind}_0 k_F^2$. This implies that materials with higher carrier densities are capable of supporting more significant spin splitting. 
 
 Importantly, we find that any spin-orbit coupling (SOC)
 produced by inversion symmetry breaking at the interface can be
 mitigated by controlling the band alignment between the two
 materials, as illustrated in Fig.\ \ref{fig1}(b). Specifically, if we
 define a band alignment parameter
 \begin{equation}
\xal=\frac{\Dv}{\DALM}\in(0,1), \qquad \DALM= \Dv+\Dc,
\label{eq:alignment}
\end{equation}
to quantify the position of the host band edge we find that there is
always a value of $x$ that results in {\em vanishing} Rashba
SOC at the Fermi surface. Thus, band alignment control can be used to transform the host into a pure
non-relativistic altermagnet.     
 Since relative band alignment in semiconductor heterostructures can be manipulated by a transverse electric field, including electrically driven transitions between distinct band-alignment regimes, demonstrated experimentally in van der Waals heterobilayers \cite{Kistner2024}, the parameter $x$ constitutes a potentially gate-tunable control knob. As we show below, this provides a direct means of electrically tuning the heterostructure toward  what we call the SOC ``sweet spot" characterized by predominantly altermagnetic spin splitting. We note that in generic systems the cancellation that leads to negligible Rashba SOC acquires a weak momentum dependence, which we characterize below.

\section{Motivation}\label{sec:motive}

Engineering a $d$-wave semiconducting altermagnet is an interesting
fundamental challenge in its own right but also has potential technological applications. Below
we briefly outline two such applications in spintronics and spin qubit
design. We also discuss broader implications of our work as it relates to general design principles for engineering non-relativistic spin splitting in semiconductors.

\subsection{Spintronic applications}

The field of spintronics seeks to leverage electron spin degree of
freedom in storing, transmitting and processing classical information
\cite{Fabian2004,Sinova2015}. Altermagnets represent an attractive
material platform for spintronic applications due to their ability to
generate and carry spin-polarized currents without producing stray
magnetic fields that cause unwanted cross-talk between the individual
circuit elements and tend to be difficult to mitigate or control.  
Spin valve, spin filter and spin splitter effects that underlie various key
spintronic functionalities require the lowest-symmetry, $d$-wave
altermagnets for optimal operation and many proposed applications also rely on
electrostatic gating to control these effects
\cite{Fu2025,Jungwirth2026b}. In sum, spintronic applications would be
best served by semiconducting $d$-wave altermagnets.

Although no single material known today falls in this class there are
several good candidates for insulating $d$-wave altermagnets. These
include a correlated layered insulator La$_2$O$_3$Mn$_2$Se$_2$
\cite{Wei2025,Asai2026,Zhang2026}, iron-based transition metal chalcogenides Fe$_2$Mo$X_4$
($X$ = S, Se, Te) and Fe$_2$WTe$_4$ \cite{Fan2025}, as well as
monolayer oxychalcogenide V$_2$Se$_2$O \cite{Ma2021}. We envision interfacing these materials with
well-characterized and commonly used non-magnetic single valley semiconductors such
as GaAs, InAs, InP, black phosphorus or wide-bandgap nitrides (GaN, InN, AlN), all of
which have a conduction band centered at the $\Gamma$ point. 
GaAs and the nitrides additionally exhibit weak intrinsic SOC which makes them suitable hosts if one is interested in engineering primarily non-relativistic spin splitting. As we
demonstrate below such a heterostructure will behave as an 
altermagnetic semiconductor with controllable spin splitting and
electron density.

\subsection{Spin qubit design}
Conventional electron-spin qubits typically rely on an applied magnetic field to generate the Zeeman splitting between their logical states. Although local electrical control can be achieved through spin-orbit coupling, magnetic-field gradients, or variations of the effective $g$-factor, these approaches introduce additional materials and device constraints. 
In addition, the applied field is often difficult to reconcile with the
presence of superconducting
resonators desired for high-fidelity readout, which tolerate stray magnetic fields
poorly. Altermagnetism offers a qualitatively different possibility: a sizable spin splitting can exist at zero applied magnetic field and, because it depends on momentum, can be controlled through the orbital confinement of the electron.
 Recently, proposals have been advanced
\cite{AbadilloUriel2026,Vakili2026}  that employ altermagnets in spin qubit
design to create \textit{field-free, electrically tunable} and 
{\em individually addressable} qubits.

When an electron in a $d$-wave altermagnet is confined to an elliptical quantum 
dot, the dot geometry selects which momenta it samples, converting this
momentum-dependent splitting into an electrically tunable
single particle spectrum with field-free spin splitting. This in turn leads to a possibility of
harnessing such quantum dots as spin qubits with  
individually addressable qubit frequency. With the net magnetization being zero,
there are no stray fields, preserving high-fidelity superconducting-resonator readout.

An important hurdle facing development of such qubits is the choice of
a suitable material. An intrinsic altermagnetic host must be, 
all at once, a $d$-wave altermagnet that is also gapped and semiconducting, single-valley with a
simple band edge, low in nuclear-spin isotopes (to suppress hyperfine dephasing), and
clean enough to control effectively electron density by electrostatic
gating. No presently known materials satisfy these 
stringent conditions.

One proposal, mentioned in
\cite{AbadilloUriel2026}, is to decouple the host from the source of
spin splitting by  proximitizing a semiconductor with an
altermagnet. In this  geometry, the confined electron lives in 
a conventional, well-optimized semiconductor, while an adjacent altermagnet imprints
its $d$-wave spin splitting across the interface via the altermagnetic proximity
effect. In this context the altermagnetic proximity effect has already been discussed
in Ref.\ \cite{Zhu2026} based on the first-principles
calculation. Here we advance an analytical theory of  this phenomenon 
by deriving the effective Hamiltonian of the proximitized
semiconductor and studying its properties. This approach helps us establish  the general conditions
under which a useful altermagnetic semiconductor can be created as
well as understand its sensitivity to perturbations such as the
relativistic SOC.

\subsection{General considerations}

For both applications it is desirable that the inherited altermagnetic splitting dominate relativistic spin-orbit effects. A sizable Rashba interaction changes the characteristic spin texture of a $d$-wave altermagnet, lifts the spin degeneracy along its nonrelativistic nodal directions, and can complicate protocols that rely on the momentum-space structure of the altermagnetic splitting. Interfaces, however, generically break inversion symmetry and therefore provide precisely such a source of Rashba coupling. The ability to suppress this interfacial contribution without reducing the altermagnetic proximity effect is consequently an important design requirement.

More broadly, the heterostructure approach separates material requirements that are difficult to satisfy simultaneously in a single compound. The semiconductor host can be selected for a simple band structure, low disorder, favorable nuclear-spin composition, and mature electrostatic control, while the magnetic donor can be optimized independently for a large $d$-wave altermagnetic splitting and an insulating gap. This separation is especially useful because the proximity-induced nonrelativistic and relativistic spin interactions need not track one another. As we show below, their distinct dependence on the host-donor band alignment makes it possible to strongly suppress the interfacial Rashba coupling without simultaneously suppressing the desired altermagnetic splitting. The heterostructure therefore provides not only a route around the scarcity of suitable intrinsic materials, but also a means of engineering the relativistic and nonrelativistic components of the spin splitting independently.

The induced altermagnetic state should moreover be directly accessible to experiment. Momentum-resolved spectroscopy of the semiconductor layer can probe its characteristic anisotropic spin splitting, while transport measurements can detect the corresponding spin-splitter response. Electrostatic control of the relative band alignment provides an additional diagnostic: tuning through the SOC sweet spot should suppress relativistic spin splitting on the altermagnetic nodal directions while leaving the $d$-wave splitting away from those directions finite.

\section{Model}\label{sec:model}

We describe the heterostructure depicted in Fig.\ \ref{fig1}(a) by a minimal continuum model
Hamiltonian consisting of a semiconductor, altermagnet and the mutual
tunnel coupling term, 
\begin{equation}
H = H_s + \HALM + H_g.
\label{eq:Htot}
\end{equation}
The semiconductor host is modeled as a low-density degenerate electron
gas with a parabolic band and no intrinsic SOC,
\begin{equation}
H_s(\kk)=\xi^s_\kk\,\sigma_0,\quad \xi^s_\kk\simeq\frac{k^2}{2m_s}.
\label{eq:Hs}
\end{equation}
Here $\sigma_a$ denotes Pauli matrices in the spin space, with
$\sigma_0$ being a unit matrix. We adopt units such that $\hbar=1$ and the lattice constant $a_0=1$, making $k$ dimensionless, and $m_s$ in the units of inverse energy.

\begin{figure}[t]
\includegraphics[width=\columnwidth]{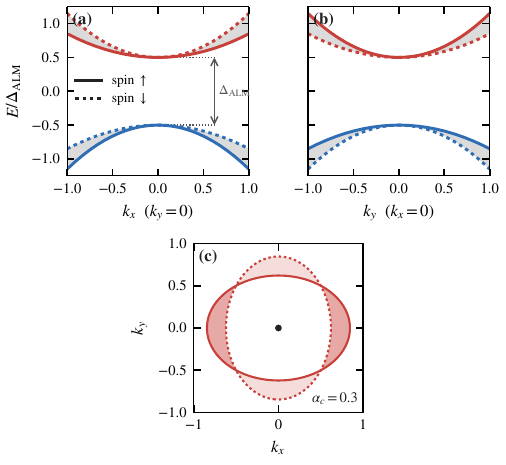}
\caption{Spectral properties of the altermagnet Hamiltonian defined in
  Eq.\   \eqref{eq:HALM}. 
(a,b)~Energy levels along $k_x$ and $k_y$: both spins are exactly
degenerate at each band edge and differ only in curvature; comparing (a) with
(b) shows the $C_4\cT$ altermagnetic symmetry. Shading marks the spin
splitting.
(c)~Constant-energy contours of the conduction band: ellipses
$90^\circ$-rotated between the spins, centered on the $\Gamma$ point
represented by the dot.
 { Valence band: blue; conduction band: red; spin $\uparrow$: solid; spin
$\downarrow$: dotted. Parameters are: $\DALM=1$~eV with $E_{c}^{\rm edge}=-E_{v}^{\rm
edge}=\DALM/2$, so that the gap is centered at zero energy,
$(m_v,m_c)=(-1,1)$~eV$^{-1}$, and $\etav=\etac=150$~meV, giving
$|\alpha_n|=0.3$, with the contour in (c) taken at
$E_c^{\rm edge}+0.25\,\DALM$.
}
}
\label{fig2}
\end{figure}

We make a key assumption that the 
donor is a gapped altermagnetic insulator and describe it using a two-band
Hamiltonian
\begin{equation}
\HALM(\kk)=\sum_{n=v,c}\Big[E_n(\kk)\,\sigma_0+\eta_n(\kk)\,\sigma_z\Big]P_n,
\label{eq:HALM}
\end{equation}
where
\begin{equation}
\eta_n(\kk)=\eta_n^{0}\left(k_y^2-k_x^2\right)
\label{eq:gammaeta}
\end{equation}
accounts for the  $d$-wave spin splitting and  $P_n$ projects onto
donor valence and conduction band $n\in\{v,c\}$.  The band energies
$E_n(\kk)$ are taken as
\begin{align}
E_{n}(\kk)=E_n^{\rm edge}+ {k^2\over 2 m_n},
\label{eq:gammamasses}
\end{align}
where $E_{n=v,c}^{\rm edge}$ denotes the position of the band edge and
$m_{n=v,c}$ are the corresponding effective masses, such that $m_v<0$
and $m_c>0$. The altermagnetic strength of each band can be  characterized by a dimensionless
anisotropy factor
\begin{equation}
\alpha_n\equiv 2 m_n\,\eta_n^{0}.
\label{eq:alphadef}
\end{equation}
Throughout we assume $|\alpha_n|<1$, so that effective inverse masses keep the sign of
$m_n$ along all directions and the constant-energy contours are ellipses, $90^\circ$-rotated
between the two spins. In addition, we will assume that 
\begin{equation}
\Delta_{\rm ALM}=E_{c}^{\rm edge}-E_{v}^{\rm  edge}>0
\label{eq:deltadef}
\end{equation}
so that the spectrum is fully gapped. Typical spin-split spectra
and constant energy contours of the Hamiltonian \eqref{eq:HALM} are
shown in Fig.\ \ref{fig2}. They exhibit the characteristic $C_4\cT$
symmetry of a $d$-wave altermagnet where $C_4$ and $\cT$ denote the
in-plane 4-fold rotation and time reversal, respectively.

Finally, we assume that the coupling between the semiconducting and
altermagnetic layers consists of spin conserving electron tunneling
parametrized by tunneling amplitudes $g_n$ and a Rashba-type spin flip
tunneling that is allowed at the interface due to the locally broken
inversion symmetry, parametrized by $\lambda_R$. This specific choice
is physically motivated: if one is interested in generating a
predominantly 
non-relativistic spin splitting in the semiconductor then, presumably,
SOC strength in the host and the donor can be minimized by a judicious choice
of materials. However, breaking of the inversion symmetry at the interface will inevitably give rise to a
non-vanishing interfacial Rashba SOC which we include in our model.     

With these assumptions the total
heterostructure Hamiltonian can be written compactly in the
second-quantized notation  as
$H=\sum_\bk\Psi_\bk^\dag\cH(\bk)\Psi_\bk$
where $\Psi_\bk=(a_{\kk\uparrow},a_{\kk\downarrow};
c_{v\kk\uparrow},c_{v\kk\downarrow},c_{c\kk\uparrow},c_{c\kk\downarrow})^T$
and
\begin{equation}
\mathcal H(\kk)=
\begin{pmatrix}
H_s(\kk) & V(\kk)\\[2pt]
V^\dagger(\kk) & \HALM(\kk)
\end{pmatrix},
\qquad
V=\begin{pmatrix} V_v, V_c \end{pmatrix}.
\label{eq:blockH}
\end{equation}
Here $a^\dag_{\bk\sigma}$ and $c^\dag_{n\bk\sigma}$ create electrons
with crystal momentum $\bk$ and spin $\sigma$ in the semiconductor
and the altermagnet, respectively. $V_n$ are $2\times 2$ matrices in spin
space describing the tunneling processes,
\begin{equation}
V_n(\kk)= g_n\,\sigma_0+\lambda_R\left(\sigma^x k_y-\sigma^y k_x\right).
\label{eq:Vgamma}
\end{equation}
Various model parameters are chosen such that the host conduction band lies
within the donor gap as illustrated in Fig.\ \ref{fig1}(b) and the
band alignment is quantified by parameter $x\in(0,1)$ defined in Eq.\ \eqref{eq:alignment}. We remark that within our conventions $\eta^0_n$ and $\lambda_R$ both have units of energy.

\section{Effective Semiconductor Hamiltonian}\label{sec:heff}

We now wish to derive the effective Hamiltonian for the proximitized
semiconductor by formally integrating out the gapped altermagnetic
degrees of freedom. This is most easily achieved by performing a
unitary Schrieffer-Wolff type transformation on the $6\times 6$ matrix Hamiltonian $\cH(\kk)$
that eliminates the off-diagonal term to leading order in the interlayer coupling. In doing this
we will follow steps outlined in Refs.\
\cite{Zhong2007,HeinsdorfFranz2026}.

\begin{figure}[t]
\includegraphics[width=\columnwidth]{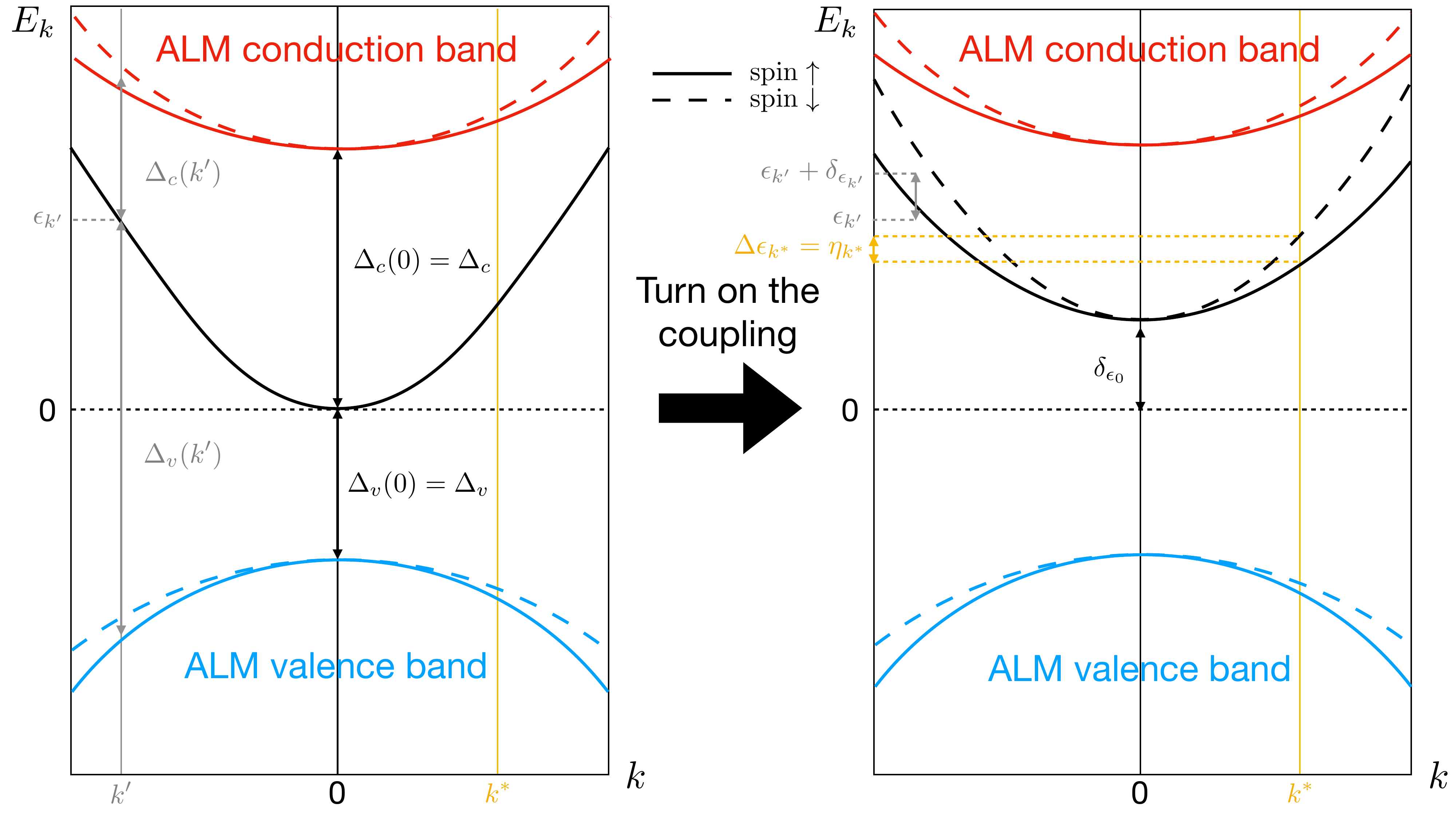}
\caption{{ Altermagnetic proximity effect. (a)~In the absence of tunneling, the donor and and the host are independent: the spin-degenerate semiconductor band $\epsilon_\kk$ (black) lies inside the gap of the altermagnet, between its spin-split valence (blue) and conduction (red) bands. It is useful to define momentum-dependent gaps $\Delta_v(\kk)$ and $\Delta_c(\kk)$ separating it from the two donor bands, measured at the same momentum. At the band edge they reduce to $\Delta_v(0)=\Delta_v$ and $\Delta_c(0)=\Delta_c$. (b)~When the tunneling is turned on, the host band undergoes a level shift $\delta\epsilon_\kk$ and acquires a spin splitting inherited from the altermagnet (spin $\uparrow$ solid, spin $\downarrow$ dashed). Because the induced couplings depend on momentum through $\Delta_n(\kk)$, the band alignment cancels the induced Rashba coupling
at one momentum only, $k^\ast$ (orange), where the splitting is purely altermagnetic, $\Delta\epsilon_{k^\ast}=\eta_{k^\ast}$. Eqs.~(\ref{eq:etaindk0},\ref{eq:lamindk0}) neglect this momentum dependence by evaluating all quantities at the band edge, $\kk=0$.}
}
\label{fig3}
\end{figure}
%

\subsection{Derivation of the effective Hamiltonian}

To reduce clutter we temporarily suppress the crystal momentum $\bk$ and write
$\mathcal H=\mathcal H_0+\mathcal V$ with 
\begin{equation}
\mathcal H_0=\begin{pmatrix} H_s & 0\\ 0 & \HALM\end{pmatrix},\qquad
\mathcal V=\begin{pmatrix} 0 & V\\ V^\dagger & 0\end{pmatrix}.
\label{eq:H0V}
\end{equation}
We next seek an anti-Hermitian generator $\cS$ such that
$\tilde{\mathcal H}=e^{\cS}\mathcal H e^{-\cS}$ is block
diagonal. Expanding,
\begin{equation}
\tilde{\mathcal H}=\mathcal H_0+\mathcal V+[\cS,\mathcal H_0]
+[\cS,\mathcal V]+\tfrac12\big[\cS,[\cS,\mathcal H_0]\big]+\dots,
\label{eq:BCH}
\end{equation}
we see that the first-order off-diagonal part vanishes if we choose
$\cS$ such that  
\begin{equation}
[\cS,\mathcal H_0]=-\,\mathcal V .
\label{eq:Scond}
\end{equation}
With the block-off-diagonal ansatz
$\cS=\left(\begin{smallmatrix}0&S\\-S^\dagger&0\end{smallmatrix}\right)$, where $S$
is a $2\times4$ block like $V$, Eq.~\eqref{eq:Scond} becomes a
linear equation
\begin{equation}
S\,\HALM-H_s\,S=-V .
\label{eq:sylvester}
\end{equation}

To solve Eq.\ \eqref{eq:sylvester}, we write out the donor structure explicitly. In the band basis,
$\HALM$ is block diagonal with one $2\times2$ spin block per band,
\begin{equation}
\HALM=\begin{pmatrix}h_v&0\\0&h_c\end{pmatrix},\qquad
h_n=E_n\,\sigma_0+\eta_n\,\sigma_z.
\label{eq:halmblocks}
\end{equation}
Consequently, $S$ and $V$ split into the corresponding column blocks
with $S=(S_v,S_c)$ alongside
$V=(V_v,V_c)$. Because $H_s$ is
proportional to the identity at fixed $\kk$, Eq.~\eqref{eq:sylvester} decouples
into one $2\times2$ equation per band,
\begin{equation}
S_n\,h_n-\xi^s_\kk\,S_n=-V_n.
\label{eq:sn}
\end{equation}
Each can be solved by right-multiplying with the inverse of the
$2\times2$ matrix
\begin{equation}
G_n=(\xi^s_\kk\sigma_0-h_n)^{-1}
\label{eq:gn}
\end{equation}
which exists as long as the level lies inside the donor
gap. Reassembling the blocks, we finally obtain
\begin{equation}
S=V\,(\xi^s_\kk(\sigma_0\otimes\sigma_0)-\HALM)^{-1}=(V_v\,G_v,V_c\,G_c).
\label{eq:sgen}
\end{equation}
We observe that each band carries its own generator
$S_n=V_nG_n$ and there are no $v$–-$c$ cross terms at this
order. Physically, this can be understood by considering that 
at second order in interlayer tunneling the relevant processes involve
virtual transitions to and from another band and hence cannot involve
both conduction and valence band simultaneously.  

Substituting Eq.~\eqref{eq:Scond} back into Eq.~\eqref{eq:BCH}, the two
second-order terms combine,
$[\cS,\mathcal V]+\tfrac12[\cS,[\cS,\mathcal
H_0]]=\tfrac12[\cS,\mathcal V]$, and the effective host Hamiltonian becomes 
\begin{equation}\label{heff}
  \Heff = H_s + \tfrac12\big([\cS,\mathcal V]\big)_{ss} = H_s + \Sigma,
\end{equation}
where $(\cdot)_{ss}$ denotes the semiconductor block. The correction
is contained in the surviving diagonal (semiconductor) block  and reads
\begin{equation}
\Sigma=V\,(\xi^s_\kk-\HALM)^{-1}\,V^\dagger ,
\label{eq:SW2}
\end{equation}
Because $\HALM$ is diagonal in the band index, the self-energy can be
expressed as  a sum over bands,
\begin{equation}
\Sigma(\kk)=\sum_{n=v,c} V_n(\kk)\,G_n(\kk)\,V_n(\kk)^\dagger,
\label{eq:channelsum}
\end{equation}
with no $v$--$c$ cross terms, and we restored the momentum
dependence. We note that the expansion is controlled by
\begin{equation}\label{eq:gmin}
  g_n\ll\min(\Dv,\Dc)=\DALM\min(\xal,1-\xal)
\end{equation}
  as each factor of $\cS$ in the
commutators is of order $g_n/\Delta_n$.

Performing explicitly the inversion indicated in Eq.\ \eqref{eq:gn}
the Green's function becomes
\begin{equation}
G_n(\kk)=\frac{\big(\xi^s_\kk-E_n(\kk)\big)\,\sigma_0
+\eta_n(\kk)\,\sigma_z}
{\big(\xi^s_\kk-E_n(\kk)\big)^2-\eta_n(\kk)^2}.
\label{eq:green}
\end{equation}
In order to make further analytical progress it is useful to adopt an
additional assumption, namely that the altermagnetic spin splitting
$\eta_n(\bk)$ is small compared to both $\Dv$ and $\Dc$. Then, it is
permissible to expand the above expression to leading order as
\begin{equation}
\begin{aligned}
G_v(\bk)&\simeq+\frac{\sigma_0}{\Dv(\kk)}
+\frac{\eta_v(\kk)}{\Dv(\kk)^{2}}\,\sigma_z, \\[2pt]
G_c(\bk)&\simeq-\frac{\sigma_0}{\Dc(\kk)}
+\frac{\eta_c(\kk)}{\Dc(\kk)^{2}}\,\sigma_z,
\end{aligned}
\label{eq:Gedge}
\end{equation}
where $\Dv(\kk)$ and $\Dc(\kk)$ are momentum-dependent gaps defined in
Fig.\ \ref{fig3}(a) that satisfy $\Delta_n(0)= \Delta_n$.

Evaluating the channel sum indicated in Eq.\ \eqref{eq:channelsum} we
finally obtain the self energy correction to the semiconductor
Hamiltonian of the form
\begin{equation}
  \Sigma(\kk)=\delta\epsilon_\bk+\eta^{\rm ind}_\bk(k_y^2-k_x^2)\sigma_z
  +\lambda^{\rm ind}_\bk(\sigma_xk_y-\sigma_yk_x).
\label{eq:sigma}
\end{equation}
Here
\begin{align}
\delta\epsilon_\kk&=\frac{\gv^2+\lambda_R^2k^2}{\Dv(\kk)}
-\frac{\gc^2+\lambda_R^2k^2}{\Dc(\kk)},
                    \label{eq:dEk}
\end{align}
represents a spin-independent correction to the semiconductor band energy, while 
\begin{align}
\eta^{\rm ind}_\kk&=\frac{\big(\gv^2-\lambda_R^2k^2\big)\,\etav}{\Dv(\kk)^{2}}
              +\frac{\big(\gc^2-\lambda_R^2k^2\big)\,\etac}{\Dc(\kk)^{2}},
                    \label{eq:etaindk}
 \end{align}
 and
\begin{align}
\lambda^{\rm ind}_\kk&=2\lambda_R\left[\frac{\gv}{\Dv(\kk)}
-\frac{\gc}{\Dc(\kk)}\right].
\label{eq:lamindk}
\end{align}
give the induced altermagnetic splitting and Rashba SOC, respectively.

\subsection{Effective Hamiltonian properties}

The spin-independent correction
$\delta\epsilon_\kk$ to the band energy is unimportant as long as it remains small compared to $\DALM$ such that the host band remains well inside the altermagnet gap as illustrated in Fig.~\ref{fig3}(b). The altermagnetic proximity effect
and the interface-induced SOC coupling in the semiconductor host are quantified by Eqs.\
(\ref{eq:etaindk},\ref{eq:lamindk}) respectively. We observe
that these receive contributions from both donor bands.  Importantly,
the band contributions to $\eta^{\rm ind}_\kk$ {\em add up} whereas
they subtract in $\lambda^{\rm ind}_\kk$. Right away we see that this
structure allows for a vanishing Rashba SOC accompanied by a large
non-relativistic spin splitting.
\begin{figure}[t]
\includegraphics[width=\columnwidth]{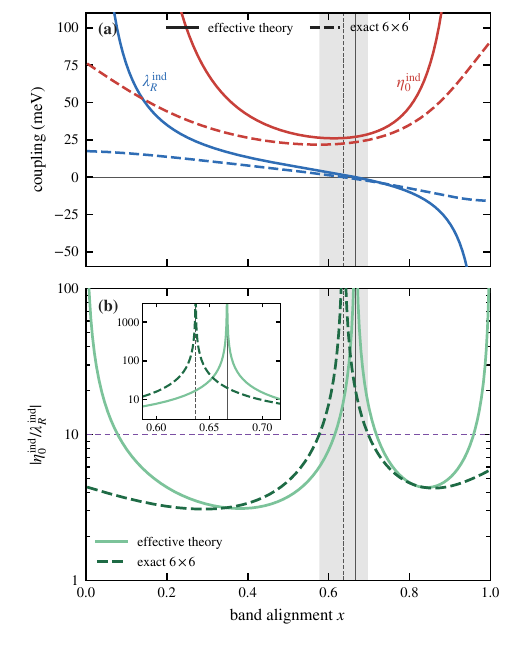}
\caption{{ Properties of the effective Hamiltonian compared to the exact solution. Solid lines represent the effective theory {Eqs.~(\ref{eq:etaindk0},\ref{eq:lamindk0})} while  dashed lines are extracted from the exact spectrum of Eq.~\eqref{eq:blockH} via Eq.~\eqref{eq:extraction} at $|\kk|=0.1$. (a)~The altermagnetic coupling $\eta^{\rm ind}_0$ (red) and the Rashba coupling $\lambda^{\rm ind}_R$ (blue). (b)~The ratio
$|\eta^{\rm ind}_0/\lambda^{\rm ind}_R|$. Two vertical lines mark the sweet spot:
$x^\ast=0.667$ from Eq.~\eqref{eq:x*} (solid) and the exact value
$x=0.637$ from Eq.~\eqref{eq:xrexactk} (dashed), at which the exact
$\lambda^{\rm ind}_R$ vanishes; the inset zooms into the two peaks. 
The shaded band highlights the region where Eq.\ \eqref{eq:criterion} is satisfied for $k_F=0.1$ and the system may be regarded as predominantly altermagnetic.
Parameters used are: $\gv=0.20$~eV, $\gc=0.10$~eV, $\etav=\etac=150$~meV,
$\lambda_R=20$~meV, $\DALM=1$~eV, $m_s=3$~eV$^{-1}$, and
$(\bar m_v,\bar m_c)=(-1,1)$~eV$^{-1}$.} 
}
\label{fig4}
\end{figure}

As the above sign structure is of key importance, it is worthwhile to pause here to understand  its physical origin. Each term in the self-energy describes a virtual transition in which the host electron tunnels into the donor band $n$ and back. The level shift $\delta\epsilon_\kk$ and the induced Rashba coupling $\lambda^{\rm{ind}}_\kk$ are energies produced by tunneling itself, so their sign is set by the energy denominator $\xi^s_\kk-E_n$, which is positive for the valence band and negative for the conduction band. This is the origin of the negative sign between contributions in Eqs.\ (\ref{eq:dEk},\ref{eq:lamindk}).  For the level shift, the transition either conserves spin in both tunneling events, giving the terms proportional to $g_n^2$, or flips it twice, giving terms proportional to $\lambda_R^2k^2$; in either case there is no net spin flip. The Rashba coupling instead requires exactly one spin flip: the spin is flipped on entering or on leaving the donor, complemented by a spin-conserving process. Hence $\lambda^{\rm ind}_\kk$ is linear in $g_n$ and the two bands again contribute with opposite signs. The altermagnetic splitting is different in kind. Neither tunneling process treats the two host spin states differently: the coupling $g_n$ has the same amplitude for a host electron of either spin, and the interface Rashba term flips host spin up to down and down to up with equal strength. Without the donor splitting, $\eta_n=0$, both host spin states would be shifted by the same amount, $(g_n^2+\lambda_R^2k^2)/\Delta_n$. The spin dependence of the host must therefore come from the donor itself, where $\eta_n(\kk)\sigma_z$ acts as a momentum-dependent exchange field. A host electron of given spin reaches the donor band either with its spin conserved, with probability $g_n^2/\Delta_n^2$, or with its spin flipped, with probability $\lambda_R^2k^2/\Delta_n^2$; in the second case it occupies the opposite spin branch and feels the exchange field with the opposite sign. Averaging gives $(g_n^2-\lambda_R^2k^2)\,\eta_n/\Delta_n^2$, the two terms of Eq.~\eqref{eq:etaindk}. Each band therefore contributes with the sign of its own $\eta_n^0$, and the two contributions to $\eta^{\rm ind}_\kk$
reinforce each other when $\etav$ and $\etac$ have the same sign. 

Importantly, symmetry alone does not fix this relative sign; whether they tend to cancel or reinforce each other depends on the microscopic structure of the altermagnet, as we show with a lattice model in Appendix~\ref{sec:microsign}. We also note that La$_2$O$_3$Mn$_2$Se$_2$ ~\cite{Wei2025} exhibits the band structure indicative of the same sign of $\etav$ and $\etac$. We thus expect the two contributions to $\eta^{\rm ind}_0$ to act in concert in this material.

To gain additional insight into the results obtained in the previous subsection it is
helpful to focus on the case of a dilute electron density in the
semiconductor, which is also the relevant limit for potential
applications.  Referring to Fig.\ \ref{fig1}(b)  we see that the Fermi
energy then lies very close to the bottom of the host band and one can
thus safely neglect all the momentum dependence entering
$\delta\epsilon_\kk$, $\eta^{\rm ind}_\kk$ and $\lambda^{\rm
  ind}_\kk$, {i.e.\ evaluate them at the band edge $\kk=0$ in Fig.~\ref{fig3}}. Focusing on the effect of the band alignment
parametrized by $x$ the effective host Hamiltonian  thus becomes
\begin{equation}
 H_{\rm eff}\simeq H_s'+\eta^{\rm ind}_0(k_y^2-k_x^2)\sigma_z
  +\lambda^{\rm ind}_R(\sigma_xk_y-\sigma_yk_x),
\label{eq:Heff0}
\end{equation}
where $H_s'=H_s+\delta\epsilon_0$ with 
\begin{align}
\delta\epsilon_0&={1\over \DALM}\left(\frac{\gv^2}{x}
-\frac{\gc^2}{1-x}\right),
                  \label{eq:dEk0}
\end{align}
                  and
\begin{align}
\eta^{\rm ind}_0&={1\over \DALM^2}\left(\frac{\gv^2 \etav}{x^2}
              +\frac{\gc^2 \etac}{(1-x)^2}\right),
                    \label{eq:etaindk0}\\
\lambda^{\rm ind}_R&={2\lambda_R\over \DALM}\left(\frac{\gv}{x}
-\frac{\gc}{1-x}\right).
                        \label{eq:lamindk0}                      
\end{align}
We observe that  $\delta\epsilon_0$ simply moves the host band relative to the
altermagnet band edge and, assuming its value is small compared to
$\DALM$, we can  ignore it henceforth.  $\eta^{\rm ind}_0$ on the
other hand is important as it transforms the host  to an
altermagnetic semiconductor. $\lambda^{\rm ind}_R$  quantifies the amount relativistic
Rashba SOC induced by the inversion symmetry breaking at the
interface. It is important to note that the above expressions are
valid only away from $x=0,1$ as per the expansion criterion given in Eq.\ \eqref{eq:gmin}.
Typical dependences of $\eta^{\rm ind}_0$ and $\lambda^{\rm ind}_R$ on the band alignment are illustrated in Fig. \ref{fig4}(a).  

Because  $\eta^{\rm ind}_0$ and $\lambda^{\rm ind}_R$ exhibit
different dependences on $x$ it is possible to use band alignment to effectively
control the host properties. For example assuming that $g_c$ and
$g_v$ are both positive one can completely eliminate Rashba SOC by
taking  $x=x^*$ in Eq.\ \eqref{eq:lamindk0} with
\begin{equation}
x^*={g_v\over g_v+g_c}.
\label{eq:x*}
\end{equation}
Inspecting Eq.\ \eqref{eq:etaindk0} it is clear that there is no
reason to expect that $\eta^{\rm ind}_0$ would vanish for this value
of $x$ -- hence a band alignment characterized by $x=x^*$ gives a pure
non-relativistic spin splitting in the host semiconductor.
We henceforth refer to this situation as the SOC ``sweet spot". In
addition, because $\lambda^{\rm ind}_R$ can have either sign, it should
be possible to achieve the same SOC-free situation even if the host Hamiltonian
itself contains a small Rashba component.

If $\etav$ and $\etac$ have the same sign then the two terms in Eq.\ \eqref{eq:etaindk0} act in concert and one may expect to obtain a large value of $\eta^{\rm ind}_0$. As mentioned above, the relative sign of $\etav$ and $\etac$ is a material specific property and can be understood based on a simple physical model.

More generally, one may be interested in engineering a semiconductor such that the electrons near the Fermi surface experience predominantly non-relativistic spin splitting.  A reasonable criterion for this is that the altermagnetic splitting at the Fermi surface 
\begin{equation}
\Delta E_{\rm ALM}\simeq 2\eta_0^{\rm ind}k_F^2
\label{eq:ALM-split}
\end{equation}
 is larger than the corresponding Rashba splitting 
\begin{equation}
\Delta E_{\rm R}\simeq 2\lambda_R^{\rm ind}k_F.
\label{eq:R-split}
\end{equation}
Requiring $\Delta E_{\rm ALM}>\Delta E_{\rm R}$ leads to the condition    
\begin{equation}
\eta_0^{\rm ind}/\lambda_R^{\rm ind}>k_F^{-1}.
\label{eq:criterion}
\end{equation}
Linear vs.\ quadratic growth with $k_F$ of $\Delta E_{\rm R}$ and $\Delta E_{\rm ALM}$ implies that the condition is easier to satisfy at higher electron densities. Figure \ref{fig4}(b) shows the ratio $\eta_0^{\rm ind}/\lambda_R^{\rm ind}$ evaluated from  Eqs.\
(\ref{eq:etaindk0},\ref{eq:lamindk0}). Assuming, as an example, that the electron density is such that $k_F=0.1$ the system will be predominantly altermagnetic for band alignments inside the shaded region centered at $x^*$. We note that for the chosen material parameters, Eq.\ \eqref{eq:criterion}  will be satisfied for all $x$ when $k_F\gtrsim 0.3$.

To get some intuition for these values we recall that in a 2D electron gas $k_F^2=2\pi n a_0^2$ where $n$ denotes the electron density and $a_0$ is the lattice constant. Taking GaN  with $a_0\simeq 0.32$ nm as an example, we find that $k_F= 0.1$ requires the electron density $n\simeq 10^{12} {\rm cm}^{-2}$ which is achievable in AlGaN/GaN heterostructures \cite{Manfra2002}. Assuming $x$ is tuned close to its sweet spot value, Fig.\ \ref{fig4}(a) gives $\eta_0^{\rm ind}\simeq 25$ meV which according to Eq.\ \eqref{eq:ALM-split} translates to $\Delta E_{\rm ALM}\simeq 0.5$ meV. This is a small splitting which may nevertheless be observable in transport experiments at low temperature. Larger electron densities will produce larger altermagnetic spin splitting.

As a consistency check on the effective Hamiltonian approach we extracted the $\eta_0^{\rm ind}/\lambda_R^{\rm ind}$ ratio
from the exact numerical solution of the full $6\times6$ Bloch Hamiltonian Eq.~\eqref{eq:blockH}, as described
{ at the end of Sec.\ \ref{sec:numerics}}. These results are plotted as dashed lines in Fig.\ \ref{fig4}. We observe that the exact result largely agrees with the approximation near the $x=x^*$ sweet spot but the two
disagree in the vicinity of $x=0,1$. The disagreement is due to the
fact that, in view of Eq.\ \eqref{eq:gmin}, the procedure leading to the effective Hamiltonian breaks down in the limit $x\to 0,1$.
The results shown in Fig.\  \ref{fig4} nevertheless indicate that, generically, there will be a range of favorable band alignments near $x=x^*$ where the ratio $\eta^{\rm
  ind}_0/\lambda^{\rm ind}_R$ can be made arbitrarily large and the
heterostructure  behaves as a pure non-relativistic altermagnet.

We note that  Fig.\ \ref{fig4} also shows small deviations between
the approximate and the exact solution near  $x=x^*$. These can be
attributed to two effects. First, the sweet spot position $x^*$ given in
Eq.\ \eqref{eq:x*} is based on approximate values of $\eta^{\rm
  ind}_0$ and $\lambda^{\rm ind}_R$ given in Eqs.\
(\ref{eq:etaindk0},\ref{eq:lamindk0})  which neglect the momentum
dependence of these quantities indicated in more accurate  Eqs.\
(\ref{eq:etaindk},\ref{eq:lamindk}). The upshot is that the precise
position of the SOC sweet spot is in general $k$-dependent: for a given band alignment the induced Rashba coupling vanishes at a single momentum only, $k^*$ in Fig.~\ref{fig3}(b), where the splitting is purely altermagnetic. Second,
these expressions measure the gaps from the bare semiconductor band
edge, whereas the proximitized level is shifted by $\delta\epsilon_0$,
Eq.\ \eqref{eq:dEk0}. For the parameters of Fig.\ \ref{fig4} this
moves the sweet spot from $x^*=0.667$ to $x=0.637$. Importantly, the approximate $x^*$
still lies inside the shaded region of Fig.\ \ref{fig4}, where the
exact ratio $\eta^{\rm ind}_0/\lambda^{\rm ind}_R$ exceeds 10. For most
practical purposes, therefore, the simple approximate expression in Eq.\
\eqref{eq:x*} should be more than adequate. For the sake of completeness we provide a more
detailed analysis that takes into account the additional
$k$-dependence and the level shift in Appendix \ref{sec:deadpoint}.

Importantly, the momentum dependence of the sweet spot is not unavoidable. As shown in Appendix A, the leading drift vanishes when in the ``symmetric limit" defined therein; 
the sweet spot location $x^*=1/2$ becomes exact and momentum-independent. Thus, suitable engineering of the band curvatures and tunneling amplitudes can realize an extended SOC-free altermagnetic regime rather than cancellation at a single momentum.

\begin{figure*}[!tb]
\includegraphics[width=\textwidth]{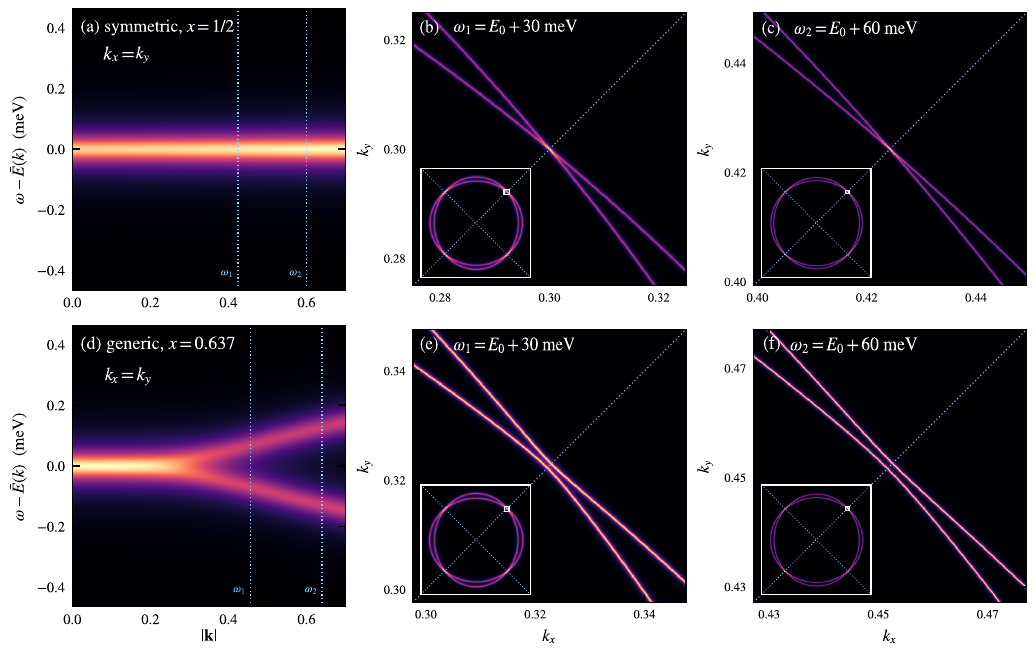}
\caption{{ Exact spectral function $A_s(\kk,\omega)$ for the semiconductor, Eq.~\eqref{eq:lorentzians}, for the symmetric (top) and generic (bottom) parameter set, each at its exact $\Gamma$-point sweet spot, Eq.~\eqref{eq:xrexact}: $\xal=1/2$ and $\xal=0.637$. (a,d)~Color plot of $A_s(\kk,\omega)$ along the nodal line $k_x=k_y$, with $\omega$ measured from the spin-averaged in-gap dispersion $\bar E(k)$. The symmetric set remains degenerate at every $|\kk|$ while for the generic set the residual $\lambda^{\rm ind}_\kk$ splits the band by $2k|\lambda^{\rm ind}_\kk|$. (b,c,e,f)~Constant-energy cuts at $\omega_1=E_0+30$~meV and $\omega_2=E_0+60$~meV, with $E_0$ the bottom of the in-gap band ($\delta\epsilon_0$ above the bare host edge), zoomed in to one of the four diagonal touching points; the insets show the full spin-split contours. We used $\delta=0.04$~meV in (a,d), $0.03$~meV in the main panel and $1$~meV in the insets. Parameters are those given in Eqs.~(\ref{eq:commonparams},\ref{eq:kindepparams}) at $\varepsilon=0.3$.
}}
\label{fig:spectraldeadpoint}
\end{figure*}

\section{Numerical results}\label{sec:numerics}

Following the strategy of Ref.\ \cite{HeinsdorfFranz2026}, we
test the effective theory against the exact spectrum: we numerically diagonalize the full
$6\times6$ Bloch Hamiltonian, Eq.~\eqref{eq:blockH}, and compute
 the electron spectral function projected onto the semiconductor block,
\begin{equation}
  A_s(\kk,\omega)=-\tfrac{1}{\pi}\Im\Tr'(\omega+i\delta-\mathcal H(\kk))^{-1},
  \label{eq:As}
\end{equation}
where
$\Tr'$ extends over the two semiconductor components. The layer-resolved
$A_s(\kk,\omega)$ is what a spectroscopic probe, e.g.\ angle-resolved
photoemission (ARPES),  of the semiconductor
would measure, and its exact in-gap dispersion accounts for the
proximity effect  non-perturbatively.

Here $\omega$ refers to the probe frequency corresponding the energy at which one asks whether the
system can absorb or emit an electron with momentum $\kk$. Scanned over a
window at fixed $\kk$, it makes $A_s(\kk,\omega)$ light up
whenever it coincides with an energy eigenvalue 
that has weight on the semiconductor. To see this, we insert the eigenbasis of
the Hermitian Bloch matrix, $\mathcal H(\kk)\ket{m,\kk}=E_m(\kk)\ket{m,\kk}$ with
$m=1,\dots,6$, into the resolvent:
\begin{equation}
\left(\omega+i\delta-\mathcal H(\kk)\right)^{-1}
=\sum_{m=1}^{6}\frac{\ket{m,\kk}\!\bra{m,\kk}}{\omega-E_m(\kk)+i\delta}.
\label{eq:resolvent}
\end{equation}
Taking the partial trace with $P_s$ the projector onto the two
semiconductor components -- in the basis defined above Eq.~\eqref{eq:blockH}, explicitly
$P_s=\mathrm{diag}(1,1,0,0,0,0)$ -- turns each numerator into the weight
$\bra{m,\kk}P_s\ket{m,\kk}=\abs{P_s\ket{m,\kk}}^2$, and
$\Im\left[1/(\omega-E_m+i\delta)\right]
=-\delta/[(\omega-E_m)^2+\delta^2]$ then gives
\begin{equation}
A_s(\kk,\omega)=\sum_{m=1}^{6}\,\abs{P_s\ket{m,\kk}}^2\,
\frac{\delta/\pi}{\left[\omega-E_m(\kk)\right]^2+\delta^2}.
\label{eq:lorentzians}
\end{equation}
Hence $A_s(\kk,\omega)$ can be interpreted as a set of Lorentzian peaks of width $\delta$ centered on the six eigenvalues
$E_m(\kk)$, each weighted by the semiconductor content
$\abs{P_s\ket{m,\kk}}^2$ of its eigenstate.

{ We evaluate the spectral function for two sets of parameters, a symmetric and a generic, at a coupling strength $\varepsilon=(\gv+\gc)/\DALM=0.3$. Both sets share
\begin{equation}
\begin{gathered}
\DALM=1~\text{eV},\qquad m_s=3~\text{eV}^{-1},\qquad
|\bar m_v|=1~\text{eV}^{-1},\\
\etav=\etac=150~\text{meV},\qquad
\lambda_R=20~\text{meV},
\end{gathered}
\label{eq:commonparams}
\end{equation}
while the tunneling amplitudes and conduction band mass follow one of two rules:
\begin{equation}
\begin{gathered}
\text{symmetric:}\quad \gv=\gc=\tfrac{1}{2}\,\varepsilon\DALM,\quad
\bar m_c=0.6~\text{eV}^{-1};\\
\text{generic:}\quad \gv=2\gc=\tfrac{2}{3}\,\varepsilon\DALM,\quad
\bar m_c=1~\text{eV}^{-1}.
\end{gathered}
\label{eq:kindepparams}
\end{equation}
}

Figure~\ref{fig:spectraldeadpoint} shows $A_s(\kk,\omega)$ for the
two parameter sets where each is placed at its sweet spot:
the alignment that gives $\lamind=0$ at $\kk=0$. Because the sweet spot generally moves with momentum, we must pick one momentum value at which we impose it; we pick the $\Gamma$, which coincides with Eq.\ \eqref{eq:x*}. For the symmetric parameter set this choice has no
effect, since, as we discuss in Appendix \ref{sec:deadpoint}, $x^\ast=1/2$ remains true at every momentum, { the band remains degenerate along the whole nodal line [Fig.~\ref{fig:spectraldeadpoint}(a)], and the two constant-energy contours cross on the diagonals [(b,c)]. For the generic parameter set, the sweet spot condition does not hold away from $\Gamma$: the residual $\lambda^{\rm ind}_\kk$ splits the band on the nodal line [(d)], and the crossing of the contours becomes an avoided crossing whose gap grows with the ring radius [(e,f)]. The effect is small, so we can say that to a good accuracy the host remains a nonrelativistic altermagnet.} 

{ The exact ratio $\eta^{\rm ind}_0/\lambda^{\rm ind}_R$ shown in
Fig.~\ref{fig4} is extracted directly from the spectrum of the full
Bloch Hamiltonian Eq.~\eqref{eq:blockH} using a procedure based on the fact that the two couplings have different angular dependences. From the effective Hamiltonian,
Eq.~\eqref{eq:sigma}, the two in-gap bands are split by
\begin{equation}
\Delta E(\kk)=2\sqrt{\left[\eta^{\rm ind}_0\left(k_y^2-k_x^2\right)\right]^2
+\left(\lambda^{\rm ind}_R\,k\right)^2}.
\label{eq:splittingmain}
\end{equation}
The altermagnetic term vanishes on the nodal line  $k_x=k_y$, while the
Rashba term depends only on $k=|\kk|$. The splitting on the nodal line,
$\Delta E_{\rm n}=2|\lambda^{\rm ind}_R|k$, therefore measures the
Rashba coupling alone. The splitting along the antinodal line $k_y=0$ is $\Delta E_{\rm a}=2\sqrt{(\eta^{\rm ind}_0k_x^2)^2+(\lambda^{\rm
ind}_Rk_x)^2}$ and can be used to extract $\eta^{\rm ind}_0$. At each band alignment
$x$ we diagonalize Eq.~\eqref{eq:blockH} at the same momentum $|\kk|$
on these two cuts, and read off the two splittings $\Delta E_{\rm n,a}$. From these we then calculate 
\begin{equation}
|\lambda^{\rm ind}_R|=\frac{\Delta E_{\rm n}}{2k},\ \
|\eta^{\rm ind}_0|=\frac{1}{k^2}
\sqrt{\left(\frac{\Delta E_{\rm a}}{2}\right)^2
-\left(\lambda^{\rm ind}_R\,k\right)^2},
\label{eq:extraction}
\end{equation}
and plot the extracted values in Fig.~\ref{fig4}.
}

\section{Conclusions}\label{sec:conclusions}
In this work, we have proposed a route to engineer a semiconducting $d$-wave altermagnet by interfacing a conventional semiconductor with an insulating altermagnetic donor. Starting from a minimal continuum model, we show that virtual interlayer tunneling transfers the momentum-dependent spin splitting of the altermagnet to the semiconductor. 
The resulting proximity-induced splitting is of the second order in the tunneling amplitude and the relevant altermagnetic strength $\eta_0^{\rm ind}$ can reach tens of meV for realistic parameters. 
The resulting spin splitting experienced by the electrons near the Fermi surface depends on the carrier density through its dependence on $k_F$ indicated in Eq.\ \eqref{eq:R-split},
suggesting that a substantial altermagnetic proximity effect can be realized at higher electron densities without requiring the semiconductor itself to possess magnetic order.

An important feature of the proposed heterostructure is the possibility of controlling the relativistic and nonrelativistic components of the induced spin splitting independently through band alignment. Inversion-symmetry breaking at the interface generically produces a Rashba spin-orbit coupling, but the virtual processes involving the valence and conduction bands of the altermagnet contribute to this coupling with opposite signs. Their contributions to the altermagnetic splitting, by contrast, can reinforce one another. This leads to an SOC sweet spot at which the induced Rashba coupling vanishes while a finite $d$-wave altermagnetic splitting remains. Exact diagonalization of the full heterostructure Hamiltonian confirms this mechanism beyond the perturbative effective theory and shows how the location of the sweet spot acquires weak momentum dependence in a generic system.

It is important to note that, the relative band alignment in semiconductor heterostructures is not fixed solely by the constituent materials, but can be controlled by an applied transverse electric field or electrostatic gating, as demonstrated experimentally in van der Waals semiconductor heterobilayers \cite{Kistner2024} and explored theoretically in a variety of heterostructures \cite{Xiong2017,Nguyen2022}. Thus, the all-important alignment parameter $x$ introduced in Eq.\ \eqref{eq:alignment} may be regarded, at least in part, as an experimentally tunable control parameter.

These results suggest that proximity engineering can circumvent an important materials constraint confronting proposed altermagnetic devices. Rather than requiring a single material to simultaneously provide a suitable semiconductor band structure, $d$-wave altermagnetism, and weak spin-orbit coupling, these functions can be divided between two materials and optimized separately. The semiconductor can provide a clean, well-characterized and gate-tunable electronic platform, while the adjacent insulating altermagnet supplies the desired nonrelativistic spin splitting. Such heterostructures could therefore provide a versatile platform for electrically controlled spintronic devices and field-free spin qubits, while also opening a broader route toward engineering altermagnetic electronic states by proximity.

Looking ahead, several research directions are suggested by the present study. It would be clearly desirable to impart a larger altermagnetic splitting to the host semiconductor. A key limiting factor in the present setup with the valley located at the $\Gamma$ point is the $k_F^2$ suppression implied by Eq.\ \eqref{eq:ALM-split}. For realistic semiconductor carrier densities this amounts to a $\sim 10-100$ fold suppression compared to the bare value of $2\eta^{\rm ind}_0$. This suppression reflects the simple fact that spin splitting in a $d$-wave altermagnet vanishes as $k^2$ near $\Gamma$. Using a semiconductor with valleys located near X and Y points of the Brillouin zone would avoid this suppression as the $d$-wave splitting behaves as $\sim(\cos{k_x}-\cos{k_y})$ and is therefore maximal at these points. Several families of materials with valleys near X and Y are known to exist; these include (001) films of rock-salt group IV-VI compounds (SnTe, PbTe, PbSe, PbS) and group-IV monochalcogenides (SnS, SnSe, GeS, GeSe). A detailed study of heterostructures with these materials would be interesting.

Our study also made several simplifying assumptions that could be relaxed in future works if one aims at obtaining more realistic and material-specific predictions. These include perfect momentum conservation for the interlayer tunneling processes, which clearly will not hold unless the materials are perfectly lattice matched, a neglect of any net exchange field generated by an uncompensated surface, the effects of the finite semiconductor layer thickness, and the material-specific symmetry constraints that may allow additional terms in the Hamiltonian beyond those considered here. Our work nevertheless points to rich new physics relevant to applications that is enabled by altermagnet-semiconductor heterostructures which will hopefully motivate future theoretical and experimental studies of these systems. 
\section*{Acknowledgments}

We are grateful to Nitin Kaushal  for illuminating discussions and correspondence. This research was supported in part by NSERC and CIFAR. { Claude (Anthropic) was used to assist with debugging parts of the numerical code, checking references, and improving the manuscript.}

\bibliography{refs}


\appendix

\onecolumngrid
\begin{center}
\bf
\end{center}
\vspace{0.5cm}
\twocolumngrid

\section{SOC sweet spot}\label{sec:deadpoint}
The SOC sweet spot is the band alignment at which the induced Rashba coupling vanishes, so that the two in-gap bands of Eq.~\eqref{eq:splittingmain} are degenerate on the nodal lines $|k_x|=|k_y|$. Setting $\lambda^{\rm ind}_R=0$ in Eq.~\eqref{eq:lamindk0} gives the sweet spot of Eq.~\eqref{eq:x*}. As mentioned in Sec.~\ref{sec:heff}, this result has two limitations. First, it holds only at $\Gamma$: the gaps $\Delta_n(\kk)$ change with momentum, so the sweet spot moves with $\kk$. Second, it measures the gaps from the bare host band edge, while the tunneling also shifts the in-gap level away from that edge
[Eq.~\eqref{eq:dEk}], so the electron sees one gap widened and the other narrowed.

\begin{figure}[t]
\includegraphics[width=\columnwidth]{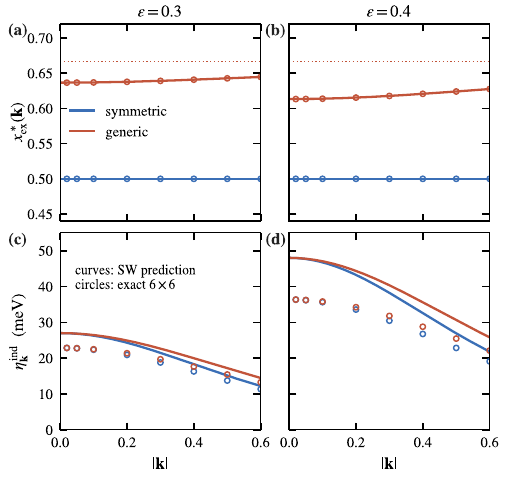}
\caption{Closed-form predictions against the exact spectrum, for the
symmetric (blue) and generic (red) sets at $\varepsilon=0.3$ (left) and
$\varepsilon=0.4$ (right). (a,b)~The sweet spot: curves are
Eq.~\eqref{eq:xrexactk}, circles are located from the $6\times6$ spectrum by
closing the nodal-line splitting at each $|\kk|$; dotted, the second-order
value $x^\ast=2/3$ of the generic set [Eq.~\eqref{eq:xrstar}]. (c,d)~The
splitting at the sweet spot: curves are Eq.~\eqref{eq:etaatxr}, circles are
the measured $\Delta E/2k^2$ along $k_x$---by Eq.~\eqref{eq:splittingmain},
with $\lambda^{\rm ind}_\kk=0$ at the sweet spot, the splitting there is
$2\eta^{\rm ind}_\kk k^2$. Parameters are: $\DALM=1$~eV, $m_s=3$~eV$^{-1}$,
$m_v=-1$~eV$^{-1}$, $\etav=\etac=150$~meV, $\lambda_R=20$~meV; symmetric
set $\gv=\gc$, $m_c=0.6$~eV$^{-1}$; generic set $\gv=2\gc$,
$m_c=1$~eV$^{-1}$; with $\gv+\gc=\varepsilon\DALM$, i.e.\
$(\gv,\gc)=(0.15,0.15)$ and $(0.2,0.1)$~eV at $\varepsilon=0.3$, and
$(0.2,0.2)$ and $(0.267,0.133)$~eV at $\varepsilon=0.4$.}
\label{fig:vstheory}
\end{figure}

In this appendix we include both effects. On the nodal lines the $d$-wave term vanishes, and the $6\times6$ Hamiltonian of Eq.~\eqref{eq:blockH} decouples exactly into two
$3\times3$ blocks, one for each spin eigenstate of the interface Rashba term. This gives the sweet spot at every momentum with the level shift included; dropping the shift recovers the second-order result. We then evaluate the induced splitting at the sweet spot and compare both with the exact spectrum in Fig.~\ref{fig:vstheory}.

\subsection{Exact sweet spot on the nodal line}\label{sec:xrexact}

\textit{Momentum-dependent gaps.}---With Eqs.~(\ref{eq:Hs},\ref{eq:gammamasses}) the
momentum-dependent gaps of Eq.~\eqref{eq:Gedge} are
\begin{equation}
\begin{gathered}
\Dv(\kk)=\Dv+a_vk^2,\qquad \Dc(\kk)=\Dc+a_ck^2,\\[2pt]
a_v=\frac{1}{2m_s}+\frac{1}{2|m_v|},\qquad
a_c=\frac{1}{2m_c}-\frac{1}{2m_s},
\end{gathered}
\label{eq:Dk}
\end{equation}
and they add up to the donor gap at the same momentum,
$\Dv(\kk)+\Dc(\kk)=\DALM(\kk)\equiv\DALM+(a_v+a_c)k^2$.

\textit{Decoupling on the nodal line.}---On $k_x=k_y$ the form factor vanishes, $\eta_n(\kk)=0$, so both donor blocks
are spin-degenerate and the only spin structure left in
Eq.~\eqref{eq:blockH} is the interface Rashba matrix
$R=\lambda_R(\sigma_xk_y-\sigma_yk_x)$, whose eigenvalues are $\pm r$ with
$r=\lambda_Rk$, since $R^2=\lambda_R^2k^2\,\sigma_0$. Let $U$ diagonalize it.
Applying the \emph{same} $U$ in all three spin spaces leaves the host block
$\xi^s_\kk\sigma_0$ and both donor blocks $E_n(\kk)\sigma_0$ untouched---each
is proportional to the identity---while turning every interface block
$V_n=g_n\sigma_0+R$ into the diagonal $2\times2$ matrix
$\mathrm{diag}(g_n+r,\,g_n-r)$. All six blocks are then diagonal in the
$\pm$ label, so the $6\times6$ splits into two independent $3\times3$
matrices, one per Rashba branch:
\begin{equation}
H_\pm=\begin{pmatrix}
\xi^s_\kk & \gv\pm r & \gc\pm r\\[2pt]
\gv\pm r & E_v(\kk) & 0\\[2pt]
\gc\pm r & 0 & E_c(\kk)
\end{pmatrix},
\label{eq:H3x3}
\end{equation}
in the basis (host, valence, conduction).

\textit{Sweet-spot condition.}---Only the host couples to the other two, so
the donor amplitudes can be eliminated exactly: for an eigenvector
$(c_s,c_v,c_c)$ at energy $E$ the last two rows give
$c_n=(g_n\pm r)\,c_s/[E-E_n(\kk)]$. Substituting these into the first row
and writing $W=E-\xi^s_\kk$ for the in-gap eigenvalue measured from the host
band, so that $E-E_v(\kk)=W+\Dv(\kk)$ and $E-E_c(\kk)=W-\Dc(\kk)$,
\begin{equation}
W=\frac{(\gv\pm r)^{2}}{W+\Dv(\kk)}-\frac{(\gc\pm r)^{2}}{\Dc(\kk)-W} .
\label{eq:secular}
\end{equation}
The sweet spot is where the two branches return the same $W$, so that the
doublet stays degenerate. Subtracting the two versions of
Eq.~\eqref{eq:secular}, the terms even in $r$ cancel and an overall $4r$
divides out, leaving
\begin{equation}
\frac{\gv}{\Dv(\kk)+W}=\frac{\gc}{\Dc(\kk)-W} .
\label{eq:deadexact}
\end{equation}
The electron does not sit at the host band edge but a distance $W$ above
it, so the gaps it sees are $\Dv(\kk)+W$ and $\Dc(\kk)-W$: one widens, the
other narrows by the same amount.

\textit{Sweet spot solution.}--- We call the common value of the two ratios in
Eq.~\eqref{eq:deadexact} $\varepsilon(\kk)$, so that
$\Dv(\kk)+W=\gv/\varepsilon(\kk)$ and $\Dc(\kk)-W=\gc/\varepsilon(\kk)$.
Adding the two removes $W$ and, with the sum rule below Eq.~\eqref{eq:Dk},
gives
\begin{equation}
\varepsilon(\kk)=\frac{\gv+\gc}{\DALM(\kk)} .
\label{eq:epsk}
\end{equation}
The level shift therefore does not change $\varepsilon$; it only changes
how $\DALM(\kk)$ is divided between the two channels, which requires $W$.
With the sweet-spot denominators, each term of Eq.~\eqref{eq:secular}
collapses, $(g_n\pm r)^2\,\varepsilon/g_n=\varepsilon(g_n\pm2r+r^2/g_n)$,
and in the difference of the two channels the $\pm2r$ pieces drop out:
\begin{equation}
W=\varepsilon(\kk)\left[(\gv-\gc)+\lambda_R^2k^2
\left(\frac{1}{\gv}-\frac{1}{\gc}\right)\right]=\delta\epsilon_\kk ,
\label{eq:Wdead}
\end{equation}
i.e.\ $W$ is the level shift of Eq.~\eqref{eq:dEk} evaluated at the sweet
spot. Then $\Dv(\kk)=\gv/\varepsilon(\kk)-\delta\epsilon_\kk$, and converting back to
the $\kk=0$ alignment fraction $\xal=\Dv/\DALM$ gives
\begin{equation}
{\;x^\ast_{\rm ex}(\kk)=\frac{\gv}{\gv+\gc}
-\frac{\gc a_v-\gv a_c}{\DALM(\gv+\gc)}\,k^2
-\frac{\delta\epsilon_\kk}{\DALM}\;}
\label{eq:xrexactk}
\end{equation}
exact for the model of Eq.~\eqref{eq:blockH} at every momentum.

\textit{Second-order limit.}---Setting $W=0$ in Eq.~\eqref{eq:deadexact}
gives back $\gv/\Dv(\kk)=\gc/\Dc(\kk)$, the condition
$\lambda^{\rm ind}_\kk=0$ of Eq.~\eqref{eq:lamindk}. Solving for the
alignment,
\begin{equation}
x^\ast(\kk)=\frac{\gv}{\gv+\gc}
-\frac{\gc a_v-\gv a_c}{\DALM\,(\gv+\gc)}\,k^2 ,
\label{eq:xrstar}
\end{equation}
which reduces to Eq.~\eqref{eq:x*} at $\Gamma$. These are the first two
terms of Eq.~\eqref{eq:xrexactk}, so the exact sweet spot is the
second-order curve lowered by the level shift in units of the donor gap. At
$\Gamma$ the drift term drops and
\begin{equation}
x^\ast_{\rm ex}=\frac{\gv}{\gv+\gc}-\frac{\delta\epsilon_0}{\DALM},\qquad
\delta\epsilon_0=\frac{\gv^{2}-\gc^{2}}{\DALM},
\label{eq:xrexact}
\end{equation}
and when $\gv=\gc$ the shift vanishes at every momentum.

\textit{Condition for $\kk$-independent sweet spot.}---Equation
\eqref{eq:xrexactk} has two sources of momentum dependence: the $k^2$ drift
and the running of the shift $\delta\epsilon_\kk$. The drift is a pure
$k^2$ term, while $\delta\epsilon_\kk$ carries all powers of $k^2$, so each
must vanish on its own. The drift vanishes when
\begin{equation}
\gc\,a_v=\gv\,a_c .
\label{eq:kindep}
\end{equation}
The shift is constant in $\kk$ only if $\gv=\gc$, since the factor
$(1-\lambda_R^2k^2/\gv\gc)/\DALM(\kk)$ in Eq.~\eqref{eq:Wdead} decreases
with $k^2$. Equation~\eqref{eq:kindep} then reads $a_v=a_c$, and
$x^\ast_{\rm ex}(\kk)=1/2$ for every $\kk$, exactly: the two channels are
indistinguishable, $\Dv(\kk)=\Dc(\kk)$ at all momenta.

In a real material $\gv\neq\gc$ in general. Equation~\eqref{eq:kindep}
alone still removes the leading drift, so we impose it in the following.
Since $a_v>0$ always, it requires $a_c>0$, i.e.\ $m_c<m_s$, and written out
in masses with Eq.~\eqref{eq:Dk} it fixes the host mass,
\begin{equation}
\frac{1}{m_s}=\frac{1}{\gv+\gc}\left(\frac{\gv}{m_c}
-\frac{\gc}{|m_v|}\right).
\label{eq:msstar}
\end{equation}

Substituting $g_n/\Delta_n(\kk)=\varepsilon(\kk)$
into Eq.~\eqref{eq:etaindk} gives
\begin{equation}
\begin{split}
\eta^{\rm ind}_\kk\big|_{\rm sweet\ spot}
=\left(\frac{\gv+\gc}{\DALM(\kk)}\right)^{\!2}
\bigg[&\left(1-\frac{\lambda_R^2k^2}{\gv^2}\right)\etav\\
+&\left(1-\frac{\lambda_R^2k^2}{\gc^2}\right)\etac\bigg].
\end{split}
\label{eq:etaatxr}
\end{equation}
The shifted gaps $\Dv(\kk)+W$ and $\Dc(\kk)-W$ of Eq.~\eqref{eq:deadexact}
have the same ratio $\varepsilon(\kk)$ and give the same result, so we
evaluate it at the exact sweet spot. At $\Gamma$ this reduces to
$\varepsilon^{2}(\etav+\etac)$.

\subsection{Comparison with the exact spectrum}\label{sec:params}

Figure~\ref{fig:vstheory} uses the symmetric and generic parameter sets in Eqs.~(\ref{eq:commonparams},\ref{eq:kindepparams}), with coupling $\varepsilon=0.3$ and as an additional example, with $\varepsilon=0.4$. All four sets satisfy Eq.~\eqref{eq:kindep} with the same host mass,
$m_s=3$~eV$^{-1}$, because Eq.~\eqref{eq:msstar} fixes $m_s$ through the
ratio $\gv/\gc$ alone: for the symmetric sets it reduces to $a_v=a_c$, and
for the generic sets the $2\!:\!1$ ratio gives the same value. Raising
$\varepsilon$ from $0.3$ to $0.4$ therefore changes only the tunneling
amplitudes, with the host and donor bands unchanged.

The curves of Fig.~\ref{fig:vstheory} show what Eqs.~\eqref{eq:xrexactk}
and \eqref{eq:etaatxr} predict for the four sets; the circles are the
exact-spectrum check, described in Sec.~\ref{sec:numerics}. Three observations from the curves:
\begin{itemize}
\item[(i)] The symmetric sweet spot does not move: $x^\ast_{\rm ex}=1/2$ at
every momentum and at either $\varepsilon$.
\item[(ii)] The generic sweet spot sits below the second-order value $2/3$ of Eq.~\eqref{eq:xrstar}. Tunneling pushes the in-gap level up by $\delta\epsilon_\kk$, so the electrons there sees a wider balance gap and a narrower conduction gap than the bare alignment suggests. To make the two channels hybridize equal again, the alignment has to move the level back toward the valence band, i.e.\ to smaller $\xal$.
\item[(iii)] In (c,d) the circles lie below the curves, and further below in (d) than in (c). The curves, Eq.~\eqref{eq:etaatxr}, give the splitting $\varepsilon(\kk)^2(\etav+\etac)$ assuming the in-gap state is entirely a host state. But that is not the case as part of it sits in the donor. The effective Hamiltonian acts on the host component
alone, and the energy of the full eigenstate instead has a weight factor caused by the donors, so the exact splitting is Eq.~\eqref{eq:etaatxr} reduced by
the host fraction of the state. A larger coupling moves more of the state
into the donor, which is why (d) deviates more than (c).
\end{itemize}

\section{Microscopic model for $\eta_{0}^{\rm ind}$}\label{sec:microsign}
\begin{figure}[!tb]
\includegraphics[width=0.96\columnwidth]{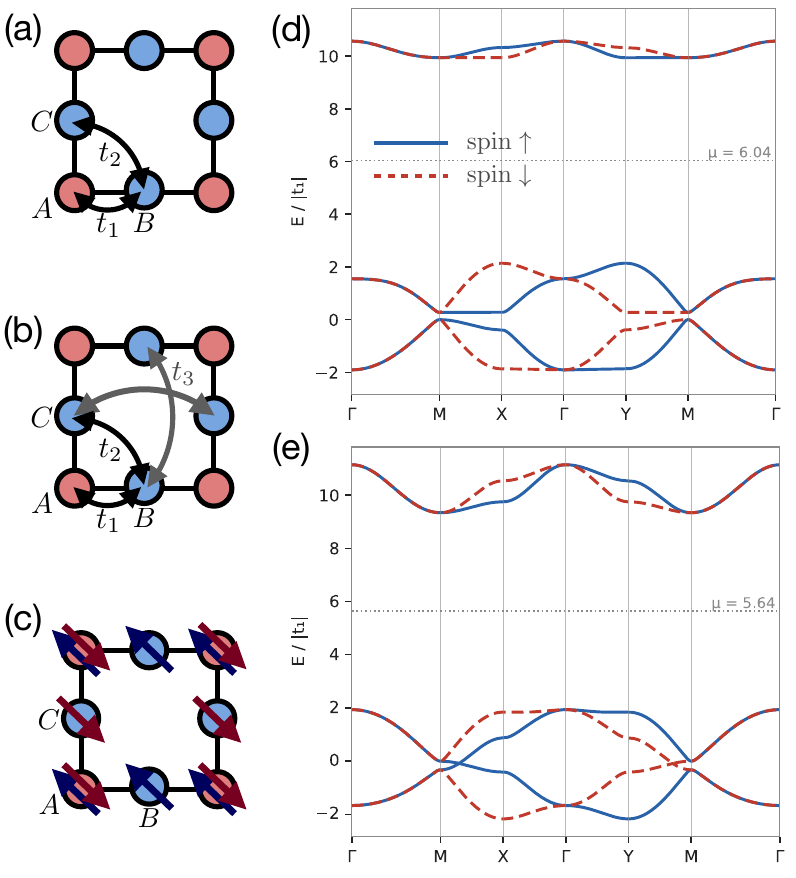}
\caption{Lattice models of the altermagnetic donor and their
Hartree--Fock bands. (a)~Modified Lieb lattice of Ref.~\cite{kaushal2024} described by 
Eq.~\eqref{eq:HLieb}: oxygen sites A (red) and transition-metal sites B and C
(blue), with hopping $t_1$ between A and B or C, and $t_2$ between B and C.
(b)~Lattice with the additional hopping $t_3$ of Eq.~\eqref{eq:H3}, B to B along
$y$ and C to C along $x$. (c)~Magnetic order of the Hartree--Fock ground
state: each A site holds two electrons with opposite
spins, the B sites carry spin up (blue arrows) and the C sites carry spin down
(red arrows). (d)~Hartree--Fock bands of Eq.~\eqref{eq:HLieb} for spin up (blue,
solid) and spin down (red, dashed); the dotted line represents the chemical
potential $\mu$.  (e)~Band structure of Eq.~\eqref{eq:H3}. Parameters used: $n=4$
electrons per unit cell, $t_1=-1$, $t_2=0.3$, $U=10$ and $\epsilon_A=0$ in
all panels, and $t_3=0.3$ in (b,e).}
\label{fig:lieb}
\end{figure}

The two terms in Eq.~\eqref{eq:etaindk0} carry the signs of $\etav$ and
$\etac$, because the weights $\gv^2/x^2$ and $\gc^2/(1-x)^2$ are positive.
When $\etav$ and $\etac$ have the same sign, the two terms add and
$\eta^{\rm ind}_0$ is nonzero at every band alignment $x$. When they have
opposite signs, $\eta^{\rm ind}_0$ vanishes at the alignment given by 
\begin{equation}
x_\eta=\frac{\gv\sqrt{|\etav|}}{\gv\sqrt{|\etav|}+\gc\sqrt{|\etac|}}.
\label{eq:xeta}
\end{equation}
For $|\etav|=|\etac|$ this point coincides with the SOC sweet spot $x^\ast$ of
Eq.~\eqref{eq:x*}, so the alignment that removes the Rashba coupling would
also remove the altermagnetic splitting. Importantly, the relative sign of $\etav$ and
$\etac$ is not fixed by symmetry. The $C_4\cT$ symmetry only requires the spin
splitting of each band to have the $d$-wave form of
Eq.~\eqref{eq:gammaeta}, but it acts on each band separately and allows
either sign of $\etav$ and of $\etac$. Thus, the relative sign is determined
by the details of the underlying microscopic  model. 

In the rest of this appendix,
we illustrate this using a specific three-band lattice model of an altermagnet relevant to  oxychalcogenides. Specifically, we show that both cases can occur and the relative sign of $\etav$ and $\etac$ can be controlled by adjusting the spatial range of the hopping between lattice sites.

We start from the modified Lieb-lattice Hubbard model of
Ref.~\cite{kaushal2024}, relevant to the transition-metal oxide layers of
oxychalcogenides such as La$_2$O$_3$Mn$_2$Se$_2$~\cite{Wei2025}. Each unit
cell has an oxygen site A and two transition-metal sites B and C
[Fig.~\ref{fig:lieb}(a)], and can be described by Hamiltonian 
\begin{equation}
\begin{split}
H={}&t_1\sum_{\langle i,j\rangle,\sigma}c^\dagger_{i\sigma}c_{j\sigma}
+t_2\sum_{\langle\langle i,j\rangle\rangle,\sigma}c^\dagger_{i\sigma}c_{j\sigma}\\
&+U\sum_{i\in\{B,C\}}n_{i\uparrow}n_{i\downarrow}
+\epsilon_A\sum_{i\in A}n_i-\mu\sum_i n_i.
\end{split}
\label{eq:HLieb}
\end{equation}
Here $t_1$ and $t_2$ follow from Fig.~\ref{fig:lieb}(a), Hubbard $U$
acts only on transition metal sites B and C, $\epsilon_A$ is the on-site energy of site A, and $\mu$ is the chemical potential. As we elaborate below, this minimal model generically gives opposite signs for $\eta^{\rm ind}$ in adjacent bands. 
We thus add a hopping between second neighbor blue sites as illustrated in Fig.~\ref{fig:lieb}(b),
\begin{equation}
H_3=t_3\sum_{j,\sigma}\left(c^\dagger_{j+\hat y,B\sigma}c_{jB\sigma}
+c^\dagger_{j+\hat x,C\sigma}c_{jC\sigma}+{\rm h.c.}\right).
\label{eq:H3}
\end{equation}
This addition maintains the original $C_4$ symmetry of the lattice and we may thus expect that $t_3$ will be non-zero in realistic situations.

We next solve the model with and without $H_3$ in collinear Hartree--Fock
theory. At electron filling $n=4$ both cases give the same altermagnetic spin order illustrated in Fig.~\ref{fig:lieb}(c). The corresponding band structures are given in Fig.~\ref{fig:lieb}(d,e). From these we can easily deduce signs of $\eta^{\rm ind}$.
In fact, since
$E_\uparrow-E_\downarrow=-2\eta_n^0k_x^2$ along $\Gamma$--X, the bands give
the signs of $\etav$ and $\etac$ directly. Without $H_3$ they are opposite, and the two contributions to $\eta^{\rm ind}_0$
cancel at $x_\eta$. On the other hand, with $H_3$ included and $t_3\gtrsim 0.1$  the conduction-band splitting reverses: $\etav$ and $\etac$ are both positive, as in
Eq.~\eqref{eq:commonparams}, and the two contributions add. 

We conclude that at the level of models, a single
hopping term that leaves the magnetic order unchanged can control the
sign structure, which therefore has to be determined for each material separately.
Inspecting the band structures for Lieb-lattice altermagnets computed in the existing literature we find that La$_2$O$_3$Mn$_2$Se$_2$ ~\cite{Wei2025} exhibit the same sign of $\etav$ and $\etac$, while iron-based transition metal chalcogenides \cite{Fan2025} and the monolayer V$_2$Se$_2$O \cite{Ma2021} exhibit opposite signs.
\clearpage
\end{document}